\documentclass[twocolumn,floats,prl,aps]{revtex4}

\usepackage{graphicx}

\newcommand\beq{\begin{equation}}
\newcommand\eeq{\end{equation}}
\newcommand\bea{\begin{eqnarray}}
\newcommand\eea{\end{eqnarray}}
\newcommand{\nonum}{\nonumber}

\begin{document}

\title{\bf Perfect Entanglement Transport in Quantum Spin Chain Systems    
}

\author{\bf Sujit Sarkar}
\address{\it 1. PoornaPrajna Institute of Scientific Research,
4 Sadashivanagar, Bangalore 5600 80, India\\
}

\date{\today}

\begin{abstract}
We propose a mechanism for perfect entanglement transport in
anti-ferromagnetic (AFM) quantum spin chain systems with modulated 
exchange coupling
along the xy plane and in the z direction.
We use the principle of 
adiabatic quantum pumping process for entanglement transfer in the spin 
chain systems. 
In our proposed mechanism, perfect
entanglement transfer can be achieved over an arbitraly long distance. 
We explain analytically and physically why the entanglement hops in alternate
sites. We
solve this problem by using the Berry phase analysis and Abelian bosonization
methods. 
We find the condition for blocking of entanglement transport even in the
perfect pumping condition. We also explain physically why entanglement transfer
in AFM chain out performs the ferromagnetic chain. 
Our analytical solution interconnects quantum
many body physics and quantum information science.   

\end{abstract}

\maketitle


{\bf 1. Introduction:}
Quantum communication between distant co-ordinates in a quantum
network is an important requirement for quantum computation
and information. One can construct the quantum network in different
ways. Optical systems typically employed in quantum communication and
cryptography application to transfer the state between two distinct
co-ordinates directly via photons \cite{kie,ski}. 
Quantum computing applications
work with
trapped atoms to transfer information between distant sites , photons
in cavity QED \cite{zeili,raus,sack,bayer,plas}. 
However we would like to study the entanglement 
transfer through
the quantum spin chain 
systems \cite{bose,chris,osbo,bayat,venuti,eckert1,eckert2,srini,hartmann,amico}. 
The equivalence of state transferring and
teleporation of information transmission has already 
been studied in the literature \cite{horo,abol}. 
The potentiality of the spin chain system, 
antiferromagnetic(AFM) and ferromagnetic(FM), as a network of quantum state and
entanglement transport has already been studied by many groups as referred in the
literature. 
The experimental evidence of nanoscale spin chain and their properties have
discussed in Ref. \cite{hein}.
Our approach in this study is different from the existing
studies in the 
literature \cite{bose,chris,osbo,bayat,venuti,eckert1,eckert2,srini,hartmann,amico}.\\
The literature of quantum entanglement study is quite vast in quantum
computation science. 
Here we mention very briefly the important works that have already existed
in state and entanglement transport in the literature: The authors
of Ref. \cite{eckert1} have shown explicitly that the quality of state and 
entanglement
transfer through all phases of spin-$1$ chain have been possible. Some
AFM phases are more efficient than the FM phase. The authors of 
Ref. \cite{eckert2} 
have shown explicitly that dimerized AFM states of spin-$1$ chains are also
able to transfer through an adiabatic modulation of exchange couplings.
The authors of Ref. \cite{venuti}, have shown explicitly that the 
quantum information
can be efficiently transferred between weakly coupled end spins of an AFM
chain because of an effective coupling between the end spins. The authors
of Ref. \cite{srini,hartmann} have studied the 
quantum state and entanglement transfer
and the authors of Ref. \cite{amico} have studied the entanglement dynamics, 
considering initial states
deviating from the final states. 
The authors
of Ref. \cite{abol,bose} have studied the entanglement transfer in a 
uniformly coupled
spin-$1/2$ AFM/FM spin chain. They have claimed a curious result that 
for the AFM spin chain, the entanglement hops to skip alternate sites. 
They have also found that the entanglement
transfer in the AFM chain outperforms the FM chain.
We explain in our
work that these theoretical predictions are natural.
Here we mention very briefly the basic mechanism
of entanglement transfer through the spin chain system based on the
conventional wisdom in the 
literature \cite{bose,chris,osbo,bayat,venuti,eckert1,eckert2,srini,
hartmann,amico}  
and at the same time illustrate
the difference with our approach.\\
It is well known that entanglement is the manifestation of quantum
correlations between two systems when they are inseparable state. We
consider the spin singlet state as an example of an entangled state.
\beq
{|{\psi}^{-} >}_{0,0'} = 
\frac{1}{\sqrt{2}}[ {|0>}_{0'} {|1 >}_{0}~-~{|1>}_{0'} { |0>}_{0} ] 
\eeq
Typically, the sender holds one member of the state of the pair of qubits while
puting the other member at the near end of the AFM spin chain of length N. The
spin chain is in the ground state.
When the spin $0$ starts to interact with the first spin of the
chain then the Hamiltonian includes this additional interaction term
( $ {I}_{0^{\prime} } {\otimes} J {{\sigma}_0} . {{\sigma}_1}$ ), 
where ${{\sigma}_0}$ and
${{\sigma}_1}$ are the Pauli spin operators for the $0$ and $1$ sites respectively
and $J$ is the exchange coupling). The
initial state being 
\beq
|{\psi} {(0)} >  =  {|{\psi}^{-} >}_{0,0'} \otimes |{ {\psi}_g } > 
\eeq 
Where $|{\psi}_{g} >$ is the ground state wave function of the AFM Hamiltonian
and $|{\psi} (0) >$ is the ground state wave function of the total Hamiltonian.
This initial state starts to evolve and from that one computes the density
matrix and concurrence to measure the entanglement and purity of states. 
But our approach
is different. 
Our main motivation is to interconnect the quantum many body physics and
quantum information science. It is common practice in quantum many body
physics to create a particle at any point in the system and study the dynamics
of that particle to understand the physical behaviour of the system.
Therefore, we consider one of the spin ($\uparrow$
or $\downarrow$) of the singlet interacts with the spin chain and this
spin itself transports through the chain medium due to the adiabatic
variation of exchange couplings of the Hamiltonian, and reaches the other
end of the chain. Our spin chains are the AFM spin chain 
with the modulated exchange couplings. 
But we consider the
monogamous nature of the shared entanglement between the two spins $0$ and $0'$.
Before we proceed further we would like to state the basic aspects 
of adiabatic
pumping process:
an adiabatic parametric quantum pump is a device that generates a
dc current by a cyclic variation of system parameters, the variation
being slow enough that the system remains close to the ground state throughout
the pumping cycle \cite{thou1,thou2}. 
It is well known that when a quantum mechanical system
evolves, it acquires a time dependent dynamical phase and time independent
geometrical phase \cite{berry}.
The geometrical phase depends on the geometry of the path
in the parameter space.
In the adiabatic entanglement pumping process, the locking potential well carries
a spin of the singlet pairs. As the locking potential well
slides through the adiabatic variation of system parameters, it
induces a current ($I$) in the system. 
In this study we 
calculate the current
of this spin transport, 
which transports a spin from one end of the chain to the other and as a 
result of which 
entanglement is transported (because the spin $0'$ and $0$ are singlet and
monogonus in nature) from one side to the other.
In our study this entanglement transport is the perfect because the
the adiabatic pumping physics based on Berry phase analysis is
topologically protected against the external 
perturbations \cite{thou1,thou2,shin}.\\ 
Here we consider two different Hamiltonian, $H_1 $ and $H_2 $ with modulated 
exchange coupling in $xy$ and $z$ directions respectively, Hamiltonians 
of the systems are the following 
\bea
{H_1} & = & - \sum_{n} J ( 1 - (-1)^{n} {{\delta}_1} (t) ) 
( {{S}_{+}}^{n}{{S}_{-}}^{n+1} +
{{S}_{+}}^{n+1}{{S}_{-}}^{n} ) \nonumber\\
& &  + \sum_{n} {\Delta} {{S}_z}^{n}{{S}_z}^{n+1} 
\eea
This model Hamiltonian has some experimental relevance \cite{shin}.
The other model Hamiltonian is
\bea
{H_2} & = & - \sum_{n} J 
( {{S}_{x}}^{n}{{S}_{x}}^{n+1} + 
{{S}_{y}}^{n}{{S}_{y}}^{n+1} ) \nonumber\\
& &  + \sum_{n} {\Delta} {{S}_z}^{n}{{S}_z}^{n+1}
- \frac{1}{2} \sum_{n} {B_0} ( 1 - (-1)^{n} {{\delta}_2} (t) ) {{S}_z}^{n}
\eea
Here we consider that the fluctuations is periodic over two lattice sites.
We see that this model have essential ingredients to capture the adiabatic
entanglement pumping.
One can express 
spin chain systems to a spinless fermion systems through 
the application of Jordan-Wigner transformation. 
In Jordan-Wigner transformation
the relation between the spin and the electron creation and annihilation 
operators
are  
$ S_n^z  =  \psi_n^{\dagger} \psi_n - 1/2 ~$, 
$ S_n^-  =   \psi_n ~\exp [i \pi \sum_{j=-\infty}^{n-1} n_j]~$, 
$ S_n^+  =  \psi_n^{\dagger} ~\exp [-i \pi \sum_{j=-\infty}^{n-1} n_j]~$,
\cite{gia2}, where $n_j = \psi_j^{\dagger} \psi_j$ is the fermion number at site $j$.
Spin operators
in terms of bosonic field are the following.
\bea
S_n^x ~&=&~ [~ c_2 \cos (2 {\sqrt {\pi K}} \phi) ~+~ (-1)^n c_3 ~]~
\cos ({\sqrt {\frac{\pi}{K}}} \theta ), \nonumber \\
S_n^y ~&=&~ -[~ c_2 \cos (2 {\sqrt {\pi K}} \phi) ~+~ (-1)^n c_3 ~]~
\sin ({\sqrt {\frac{\pi}{K}}} \theta ), \nonumber \\
S_n^z ~&=&~ {\sqrt {\frac{\pi}{K}}} ~\partial_x \phi ~+~ (-1)^n c_1
\cos (2 {\sqrt {\pi K}} \phi ) ~,
\label{spin}
\eea
\beq
 {{\psi}_{r}} (x)~=~~\frac{U_r}{\sqrt{2 \pi \alpha}}~
~e^{-i ~(r \phi (x)~-~ \theta (x))} 
\eeq
$r$ denotes the chirality of the fermionic fields,
 right (1) or left movers (-1).
The operators $U_r$ are operators that commute with the bosonic field. 
$U_r$ of different species
commute and $U_r$ of the same species anticommute. $\phi$ field corresponds to the 
quantum fluctuations (bosonic) of spin and $\theta$ is the dual field of $\phi$. 
They are
related by this relation $ {\phi}_{R}~=~~ \theta ~-~ \phi$ and  $ {\phi}_{L}~=~~ \theta ~+~ \phi$.

Using the standard machinery of continuum field theory \cite{gia2}, 
we finally get the bosonized Hamiltonians
as 
$H_{0}$ is the gapless Tomonoga-Luttinger liquid part of the Hamiltonian.
\\
After the application of continuum field-theory the Hamiltonian become, in terms
of bosonic fields.
\bea
{H_1} & = & {H_0} 
+ \frac{{E_{J_0}} {{\delta}_1}(t) }{2 {{\pi}^2} {\alpha}^2} 
\int dx : cos[2 \sqrt{K} {\phi} (x)]: \nonum\\ 
& & + \frac{\Delta}{2 {{\pi}^2} {\alpha}^2} \int dx 
: cos[4 \sqrt{K} {\phi} (x)]: 
\eea
\bea
{H_2} & = & {H_0} 
+ \frac{{B_0} {{\delta}_2}(t) }{2 {\pi} {\alpha}} 
\int dx : cos[2 \sqrt{K} {\phi} (x)]: \nonum\\
& & + \frac{\Delta}{2 {{\pi}^2} {\alpha}^2} \int dx 
: cos[4 \sqrt{K} {\phi} (x)]: 
- \frac{B_0}{2} \int dx {{\partial}_x} {\phi}(x)
\eea
Here, we would like to explain the
basic aspects of quantum entanglement pumping in terms of
spin pumping physics of our model Hamiltonians:
An adiabatic sliding motion of one dimensional potential,
in gapped Fermi surface (insulating state), pumps an integer numbers
of particle
per cycle.
In our case the transport of Jordan-Wigner
fermions (spinless  fermions) is nothing but the transport of spin from one end
of the chain to the other end because the number operator of spinless fermions
is related to the z-component of spin density \cite{cal}.
We see that non-zero ${{\delta}_1} (t)$ and ${{\delta}_2} (t)$ 
introduce the gap at
around the
Fermi point and the system is in the insulating state (Peierls insulator).
In this phase spinless fermions form the bonding orbital between the
neighboring sites, which yields a valance band in the momentum space.
It is well known that the physical behavior of the system is identical
at these two Fermi points.
We would like to analyse these double
degeneracy point,
following the seminal paper of Berry \cite{berry}:
in our model Hamiltonian there are two adiabatic parameters
${{\delta}_1} (t) $ and ${{\delta}_2}(t) $. The Hamiltonian starts to
evolve under the variation of these two adiabatic parameters, when
the Hamiltonian returns to its original form after a time
$ T$, the total geometric phase acquired by the system is
${{\gamma}_n} (T) ~= ~\frac{i}{2 \pi} \int_C
<{{\psi}_n}|{{\nabla}_R}| {\psi}_{n} >~ dR$,
a line
integral around a closed loop in two dimensional parameter
space. Using Stokes theorem, one can write
${{\gamma}_n} (T) ~= ~\frac{i}{2 \pi} \int {{\nabla}_R} \times
<{{\psi}_n}|{{\nabla}_R}| {\psi}_{n} >~ dS$.
The flux $\Phi$ through a closed surface C is, $\Phi = \int B.dS $.
Therefore one can think of the Berry phase as flux of a magnetic
field. Now we express,
${B_n} (K1) = {{\nabla}_{K1}} \times {A_n} (K1)$, and
${A_n} (K1) = \frac{i}{2 \pi} <n (K1)| {{\nabla}_{K1}} | n(K1)>$,
where $K1 = (k, {\delta}_1 (t), {\delta}_2 (t) )$.
Here $B_n$ and $A_n$ are the fictitious magnetic field (flux) and
vector potential of the
nth Bloch band respectively.
The degenerate points behave as a magnetic monopole in the
generalized momentum space ($K_1 $) \cite{berry}, 
whose magnetic unit can be shown to be
$1$, analytically \cite{shin,berry}
\beq
\int_{S1} ~dS \cdot B_{\pm} ~=~ \pm 1
\eeq
positive and negative signs of the above equations are respectively
for the conduction and valance band
meet at the degeneracy points.
$S_1 $ represent an arbitrary closed surface which enclose the
degeneracy point.
In the adiabatic process the parameter ${{\delta}_1} (t)$ or ${{\delta}_2} (t)$ 
are changed
along a loop ($\Gamma$) enclosing the origin (minima of the system).
We define the expression for spin current ($I$) from the analysis
of Berry phase.
It is well known in the literature of adiabatic quantum pumping physics
that two independent parameters are needed to achieve the adiabatic
quantum pumping
in a system \cite{sharma}. Here one may consider these two parameters as 
the real and
imaginary part of the fourier transform of a modulated coupling induce potential.
When the shape of the 
potential will change in time, then it amounts to changing the phase and
amplitude in time. The role of adiabatic parameters are not explicit
in our study. Our formalism is different from others. 
We define the expression for spin current ($I$) from the analysis
of Berry phase.
Then according to the original
idea of quantum adiabatic particle transport \cite{thou1,thou2,shin,avron},
the total number of spinless fermions ($I$)
which are transported from one side of this system to the other is equal to the
total flux of the valance band, which penetrates the 2D closed sphere
($S_2 $) spanned by
the $\Gamma$ and Brillioun zone \cite{shin}.
\beq
 I = \int_{S_2} dS \cdot B_{+1} ~=1
\eeq
$ B_{+1} $ is directly related with the Berry phase (${{\gamma}_n} (T)$)
which is
acquired by the system during the adiabatic variation of the exchange couplings
the time period of the adiabatic process.
This quantization is
topologically protected against the other perturbation as long as
the gap along the
loop remains finite \cite{shin,avron}.
Therefore the adiabatic entanglement pumping is constant over the
arbitrarily long distance of the system. This result is in contrast with 
the existed results in the literature [8,19]. 
They have found that the entanglement
decay exponentially after a certain distance.\\
Now we explain the quantum entanglement transfer for $ H_1 $.  
The second term of the Hamiltonian for NN exchange interaction has
originated from the $x$ and $y$ component of exchange interaction. 
This term implies that infinitesimal variation of 
coupling in lattice sites, is sufficient to produce a 
gap around the
Fermi points.
So when ${1/2} < K < 1$, only these 
time dependent exchange couplings 
are relevant and lock the phase operator at 
${\phi} = 0 + \frac{n \pi}{\sqrt{K}}$.
Now the locking potential slides adiabatically.
The speed of the sliding potential is low
enough such that the system stays in the same valley, i.e., 
there is no scope to jump
onto the other valley.  
The system will acquire $2 \pi$ phase during one
complete cycle of  
adiabatic process. 
This
expection is easily verified when we notice the physical meaning of the phase
operator ($\phi$ (x)). Since the spatial derivative of the phase operator
corresponds to the z-component of spin density,
this phase operator is
nothing but the minus of the spatial polarization of the z-component of
spin, i.e.,
$ P_{s^z}~= - \frac{1}{N} \sum_{j=1}^{N} j {S_j}^z $. Shindou
has shown explicitly
the equivalence between these two considerations \cite{shin}. During the
adiabatic process $ < {\phi}_{t} >$ changes monotonically and acquires
- $2 \pi$ phase. In this process $ {P_s}^{z} $ increases by 1 per cycle.
We define it analytically as
\beq
{\delta} {P_s}^{z} = \int_{\Gamma} d {P_s}^{z}
= - \frac{1}{2 \pi} \int dx {{\partial}_x} <{\phi} (x) > = 1
\eeq
This physics always hold as far as the system is locked by the sliding
potential and ${\Delta} < 1$ \cite{shin}.
The change of the spatial polarization by unity during a complete
evaluation of adiabatic cycle implies that the transport of entanglement 
across the system. This is because the spatial derivative of the
phase operator is the Cooper pair density in our system.
The entanglement transport of this scenario can be generalized up to the
value of $\Delta$  for which
$K$ is greater than 1/2
. In this limit, z-component
of the exchange interaction
has no effect on the entanglement pumping
of our system.
But when $K<1/2 $
, then the interaction due to $ \Delta $ becomes relevant and creates
a gap in the excitation spectrum. This potential profile is static. Therefore
there is no scope to slide the potential and to get a adiabatic pumping
across the system.
The authors of Ref. \cite{abol,bose} have also found that
when ${\Delta}> 1$
for $XXZ$ AFM spin chain, the fidelity of AFM spin chain also decreases
,i.e., the entanglement transport decreases in this limit.\\
Similarly for the Hamiltonian $H_2 $, the second term of the
Hamiltonian produce the gap and the pumping process is the same
as that of $H_ 1$. Therefore we conclude that the modulations in
the in plane exchange coupling and also for the modulations in the 
z-directions yield the
same adiabatic entanglement pumping.\\  
In this pumping process the most favourable states of the system
are the antiferromagnetic configuration $ |010101....>$ and
$ |101010,,,,> $ ($0$ stands for up spin and
$1$ stands for down spin). One may start from any antiferromagnetic
states and transfer the spin of every site to the right by
two sites to achieve the pumping. Therefore our test spin which
we introduce at the one end of the spin, it hops to the
right by two sites in every step. Thus when we study the entanglement
transport between the spin $0'$ and $0$, then it is natural that
entanglement also is transported through every alternate sites.
The authors of Ref. \cite{abol,bose} have observed a very peculiar
behaviour of entanglement transfer for AFM: 
the nonanalytical behaviour as a function of time. It is zero
for most of the time and it suddenly grows up and forms a peak
at a regular interval of time. But in our study the entanglement
current is constant and it is almost perfect
entanglement pumping. In their case the spin chain has the spin
rotational symmetry. When one member of an
entangled pair of qubits is transmitted through such a channel
, then the two qubits states evolve to a Werner state \cite{benn}.
But our spin chain systems there is no spin rotational invariant
symmetry and the transport mechanism is also different.   
The physical scenario of our study is completely
different from the existing physical picture.
The quantized entanglement transport of this scenario can be generalized up to the 
value of $\Delta$  for which 
$K$ is greater than 1/2. 
In this limit, the z-component
of the exchange interaction  
has no effect on the entanglement pumping
physics of  
Hamiltonian.  
. In this limit, z-component
of the exchange interaction 
has no effect on the entanglement pumping
of our system. 
\\ 
Here, we would like
to explain the difference of entanglement transport between the FM and
AFM spin chain, it has mentioned in the literature but the complete
physical explanation is not upto the 
mark \cite{bose,chris,osbo,bayat,venuti,eckert1,eckert2,srini,hartmann,amico,abol}. 
As we know that entanglement is a quantum mechanical
property, Schrodinger singled out many decades ago as "the characteristic
of quantum mechanics \cite{sch} and that has been studied extensively
in connection with Bell's inequality \cite{bell}. FM ground state 
state there is no difference between the
classical and quantum mechanical ground state and the low lying
excitations are spin-1 magnons. The AFM ground state has a complex
structure specified by the Bethe-ansatz solution. There are no similarities
between classical and quantum mechanical ground state and first excited 
state of the AFM chain and as a result of the quantum mechanical property
of the system the entanglement manifests prominently in the AFM spin chain.
This is the only clear reason why AFM outperforms the FM spin chain.\\   
Here we discuss possible sources of imperfections
in the entanglement pumping process. The non-adiabatic contributions
leave the system in an unknown superposition of states after the full
cycle. Also the appearance of Landau-Zener transition in the
pumping system should be negligible so that the system is in the
ground state. This condition limits the pumping rate of entanglement
by the mathematical relation $ \frac{h}{\tau} << J $. However even
then the entanglement pumping is not perfect due to the non vanishing
$ \frac{J}{\Delta}$. Our effort also should take the elimination of
entanglement pumping in the wrong directions. 
The residual exchange coupling may lead to
a different spin state. An entangled spin transported through a
correct exchange coupling modulation with probability $P$ and through
the residual exchange coupling with the probability $Q= 1-P$. Therefore
the pumping error in each site is $\frac{P}{Q}$. Our system consists
of $N$ sites. Therefore the probability of correct entanglement transport is
$\sim {P^{N/2}} $ and wrong entanglement transport is $\sim {Q^{N/2}}$.
The total pumping error, $({\frac{Q}{P})}^{N/2}$, decreases 
with the number of sites in nanoscale spin chain. Therefore for the 
spin chain system entanglement transport is better for larger length 
compare to the smaller length 
with same exchange couplings.\\
{\bf Conclusions:} we have presented the theoretical explanation of 
adiabatic entanglement pumping 
for our model Hamiltonians. We have found the perfect entanglement
transport condition which cure the existed results in the literature. 
We have explained few physical findings of entanglement transport which
were curious before this study.
\\
The author would like to thank, The Center for Condensed Matter 
Theory of IISc for extended facility.
Finally the author would like to thank Prof. R. Srikanth, Dr. T. Tulsi and 
Prof. Indrani Bose.

\end{document}